\documentclass[20pt,balancelastpage,nofootinbib,twocolumn,fleqn,showkeys]{revtex4-1}
\usepackage{amsfonts}
\usepackage{amsmath}
\usepackage{amssymb}
\usepackage[colorlinks, citecolor=blue,linkcolor=blue
]{hyperref}
\hypersetup{pdftitle={BEC NMS}, pdfnewwindow=true}
\usepackage{color}
\usepackage{float}
\usepackage{textcomp}
\usepackage{subfigure}
\usepackage{graphicx}
\begin{document}
\title{\textbf{Cavity Optomechanics with a Bose-Einstein Condensate: Normal Mode Splitting}} 
\author{Muhammad Asjad}
\email{asjad\_qau@yahoo.com}
\affiliation{Department of Electronics, Quaid-i-Azam University, 45320 \ Islamabad, Pakistan.}
\begin{abstract}
We study the normal mode splitting in a system consisting of a Bose-Einstein condensates (BEC) trapped inside a Fabry-P\'erot cavity driven by single mode laser field. We analyze the variations in frequency and damping rate of collective density excitation of BEC imparted by optical field. We study the occurrence of normal mode splitting which appears as consequences of hybridization of the fluctuations of intracavity field and condensate mode. It is shown that normal mode splitting vanishes for weak coupling between condensate mode and intracavity field. Moreover, we investigate the normal mode splitting in the transmission spectrum of cavity field.       
\end{abstract}
\keywords{Bose-Einstein condensate, Optomechanics, Normal mode splitting}\maketitle 
\section{introduction}
Nano optomechanical systems that couple optical degree of freedom to the mechanical motion of a cantilever have been extensively investigated in recent past \cite{Kippenberg}. In such systems, coupling is obtained via radiation pressure inside a cavity \cite{Braginsky, Mancini, Zhang}, and indirectly via quantum dots \cite{Tian} or ions \cite{Naik}. Significant progress has been made in the investigation of optomechanics, such as squeezing \cite{Fabre, Tombesi}, ultrahigh precision displacement
detection \cite{Caves,LaHaye, Rugar}, mass detection \cite{Ekinci}, gravational wave detection \cite{Caves1980,V. Braginsky}, and the transition between classical and quantum behaviour of a mechanical system \cite{Marshall}. Moreover, entangling the electromagnetic field with motional degree of freedom of mechanical systems have been explored in various approaches \cite{Vitali,Paternostro}. The entanglement generated in such systems is significant both philosophically as well as technically in relation with quantum informatics \cite{Nielsen}. Further advances in experimental techniques make possible to couple the mechanical resonators with the statistical ensemble of atoms. In such systems, the interaction is mediated by the field inside the cavity that couples the mechanical resonators to the internal and motional degrees of freedom of the atoms \cite{Ian,genes,Meiser,Ritsch}.
\paragraph*{}
The normal mode splitting (NMS) is the coupling of two degenerate modes with energy exchange taking place on a time scale faster then the decoherence rate. Moreover, the NMS is a phenomena ubiquitous in both the classical and quantum physics. The NMS, in the transmission spectrum of the cavity filed, has been observed when atoms are coupled to cavity field \cite{Thompson}. Moreover, the NMS has also been studied with artificial atoms in circuit quantum electrodynamics (QED) \cite{Wallraff} as well as in single quantum dot cavity QED \cite{Reithmaier}. In this paper, we study the occurrence of the NMS through a new optomechanical system, i.e, a coupled BEC-cavity configuration \cite{Esteve,Brennecke,Ritter,Colombe,Purdy,Baumann,Kónya,Nagy, asjad}. Briefly, the system consists of Fabry-P\'{e}rot cavity containing a condensate that effectively behaves like a vibrating mirror \cite{Ritter}. In such systems, the collective density excitation of the BEC serves the analogy of a moving mirror coupled to cavity field via radiation pressure force. The strong coupling of the quantized cavity field with the collective oscillations of the BEC has been experimentally analyzed with all the atoms are being in the same motional state \cite{Brennecke,Ritter}. Our study reveals that the frequency and the decoherence rate are sensitive to radiation pressure force. We further analyze the occurrence of NMS in the position spectrum of the condensate mode. It is observed that the position spectrum of the condensate mode splits into two peaks when coupling between condensate mode and intracavity field is considered. In addition, we also discuss the NMS in the fluctuations spectrum of the light field emitted by the cavity. We show that the distance between two peaks increases linearly with BEC-cavity field interaction. The paper is structured as follows: In Section 2, we give the theoretical model of the system and its coupling with the environment by using the Heisenberg-Langevin equations. In section 3, we solve the dynamics in frequency domain and discuss the results. Finally, we provide concluding remarks in section 4.
\section{Model and Hamiltonian of the system}
\begin{figure}[tb]
    \centering
\includegraphics[height=1in, width=3.35in]{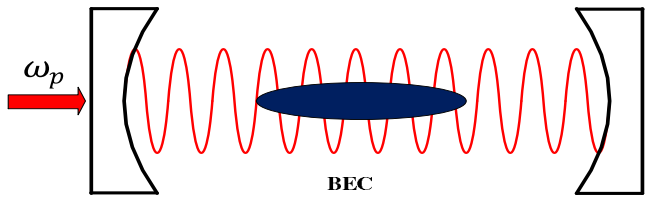}
\caption{(Color online) A sample of two-level bosonic atoms with resonant frequency $\omega_{tr}$ trapped inside a Fabry-P\'{e}rot cavity of length $L$ is interacting with standing laser field. Here, the left-end mirror is transmissive while the right-end mirror is perfectly reflecting. Moreover, the cavity is being driven by a laser of frequency $\omega_{p}$.}
\label{system}
\end{figure}

We consider an ensemble of $N$ two level bosonic atoms with resonant frequency $\omega_{tr}$ is trapped inside a Fabry-P\'{e}rot cavity and interact with standing-wave light field as shown in Fig.\ref{system}.  
We assume the atom-field detuning $\Delta_a$ is very large, therefore, one can adiabatically eliminate the excited atomic level. In the rotating frame at the driving field frequency $\omega_P$, the Hamiltonian of the Bose-Einstein condensate in the limit of small atom-atom interaction and small value of the external potential can be written as follows \cite{Ritter}:
\begin{equation}
H=\hbar\,\Delta_c c^\dag c + \hbar\,\omega_a a^\dag a + \dfrac{1}{\sqrt{2}}\hbar g \,(a^\dag + a) c^\dag c -i\hbar E(c-c^\dag) \label{ham}
\end{equation} 
where $\omega_a=4\omega_r$, $\Delta_{c}=\omega_{c}-\omega_{P}+NU_o/2$, $U_o=g^2_o/\Delta_a$, $\Delta_{a}=\omega_{P}-\omega_{tr}$, and $g=\sqrt{N}U_o/2$. Here, $N$ is the number of BEC atoms, $\omega_{tr}$ is the transition frequency and $g_o$ accounts for the coupling strength between the single intracavity photon and single condensate atom. The first term in the Eq.(\ref{ham}) stands for the energy of the intracavity mode with creation (annihilation) operator $c\,(c^\dag)$ and frequency $\omega_c$. Moreover, the empty cavity resonance frequency $\omega_{c}$ is shifted due to the presence of the BEC inside the cavity by an amount of $NU_o/2$. The second term describes the energy of the Bogoliubov mode of the collective oscillations of the BEC. Furthermore, $a(a^\dag)$ and  $\omega_{a}$ are, respectively, the creation (annihilation) operators and frequency of the condensate mode. The dynamics of the BEC can be explained as follows: The zero momentum state is only coupled to the symmetric momentum states, $\pm2\hbar k$, due to the absorption and stimulated emission of the cavity photons \cite{Ritter}. This can be explained as the condensate mode oscillates at frequency $\omega_a=4\omega_r=\hbar k^2/2m$, where $m$ is the mass of the atom and $\omega_{r}$ is recoil frequency. The third term describes the interaction between intracavity field with condensate mode. Moreover, $g$ accounts for the strength of interaction between field and BEC and it is clear from the expression of the $g$ that the single atom-photon coupling is increased by the square root of the number of condensate atoms. The last term  corresponds to the coupling between cavity mode and input laser field with coupling strength E and it is related to the input power $P$ with $|E|=\sqrt{2\kappa P/\hbar\omega_p}$, where $\kappa$ is the decay rate of the cavity field.
\paragraph*{}
 In order to describe the complete dynamics of the subsystems involved in this problem, an adequate choice is to use the formalism of the quantum Langevin equations. According to the Heisenberg-Langevin equation of motion, the commutation relations $[a,a^\dag]=1$ and $[c,c^\dag]=1$, the time evolution of the $c$ and $a$, can be obtained. We derive the Heisenberg-Langevin equations for canonical variables and introduce the noise operators $c_{in}(t)$ and $a_{in}$ weighted with the rates $\kappa$ and $\gamma$ which describe the dissipation of the intracavity field and condensate mode of collective oscillations of the BEC respectively. Therefore, Heisenberg-Langevin equations for condensate mode and intracavity field are given by    
 \begin{eqnarray}
\dot{a}&=&-(i\omega_{a}+\gamma)a-i\dfrac{g}{\sqrt{2}}\,c^\dag c+\sqrt{2\gamma}\,a_{in},\nonumber\\
\dot{c}&=& -(\kappa+i\Delta_c)\,c-i\dfrac{g}{\sqrt{2}}\,(a + a^\dag)c + E\nonumber\\ 
&&+\sqrt{2\kappa}c_{in}.\label{hle}
 \end{eqnarray} 
We assume that the noise associated with the light field is uncorrelated with the noise accounts for the condensate mode. For laser field, the noise and damping are due to the vacuum noise, losses from the cavity and fluctuations of the laser field. Moreover, the noise and damping accounts for condensate mode are due to the condensate atoms with additional nearby non-condensed atoms. In addition, $c_{in}$ and $a_{in}$ are the non-commuting noise operators associated with optical field and condensate mode respectively. They have zero mean values and nonzero correlation functions \cite{cw}:
\begin{eqnarray}
\langle\partial a^\dag_{in}(t)\partial a_{in}(t')\rangle&=&n_a \delta(t-t'), \nonumber\\
\langle\partial a_{in}(t)\partial a^\dag_{in}(t')\rangle&=&(n_a+1) \delta(t-t'), \nonumber\\
\langle\partial c^\dag_{in}(t)\partial c_{in}(t')\rangle&=&n_c \delta(t-t'), \nonumber\\
\langle\partial c_{in}(t)\partial c^\dag_{in}(t')\rangle&=&(n_c+1)\delta(t-t'), \nonumber\\
\end{eqnarray} 
where $n_c$ and $n_a$ are the occupation numbers of the optical and condensate modes, respectively. Moreover, all other correlations are zero. As we are in optical regime and a BEC at a temperature of at most a few $\mu\mathrm{K}$, therefore one can take $n_{c,a}\rightarrow 0$.    
\section{Dynamics of small fluctuations: Normal mode splitting} 
In the following we linearized the operators in Eq.(\ref{hle}) around the steady state values, $a=\left<a\right>_{ss}+\partial a$, $c=\left<c\right>_{ss}+\partial \mathrm{c}$. Here, we have assumed that the fluctuation operators $\partial a$ and $\partial c$ have zero mean. The steady state value of the intracavity mode is $\left<c\right>_{ss}=\mathrm{E}/(\kappa+i\,\Delta)$, where the total effective detuning is 
\begin{equation}
\Delta=\Delta_{c}-\dfrac{\,\omega_a g^2_{ac}}{\omega^2_{a}+\gamma^2}\,\left< c\right>_{\mathrm{ss}}. \label{detuning}
\end{equation} 
For the sake of simplicity we assume that the field is real positive and this can be achieved by adjusting the phase of the laser field. Similarly, the steady state value of the condensate mode is $\left<a\,\right>_{ss}=[-i\,g_{ac}/\sqrt{2}\,(\gamma + i\omega_{a}]\left<c\,\right>_{ss}$.
  We linearize the Langevin equations of motion given in Eq.(\ref{hle}), and assume that pump field is intense and keep terms only up to first order in the fluctuation operators. We rewrite each Heisenberg operator in Eq.(\ref{hle}) as a sum of steady state value and fluctuation operator with zero mean value. Therefore, the linearized Heisenberg-Langevin equations are,
\begin{eqnarray}
 \partial\dot{a} &= -(i\omega_{a}+\gamma)\partial a-i\dfrac{G}{2}\left(\partial c+\partial{c}^\dag\right)+\sqrt{2\gamma}\,a_{in},\nonumber \\
 \partial\dot{c} &= -\left(i\,\Delta+\kappa \right)\partial c - i\dfrac{G}{2}\left(\partial\mathrm{a}+\partial\mathrm{a}^\dag\right)+\sqrt{2\,\kappa}\, c_{in}. \label{linrz}
\end{eqnarray}
The linearized quantum Langevin equations show the fluctuations of the Bogoliubov mode as the collective oscillation of the BEC. The condensate mode is now coupled to the cavity field quadrature fluctuations by the effective couplings $G=\sqrt{2}\,g c_s$ which can be made very large by increasing the amplitude $c_s$ of the intracavity field. Linearized quantum Langevin equations (\ref{linrz}) and their corresponding Hermitian conjugate form a system of four first-order coupled operator equations, for which the Ruth-Hurwitz criteria \cite{Hurwitz} implies that the system will be stable only if the following stability conditions are satisfied,
\begin{eqnarray}
S_1&=&2\kappa\gamma\{[\kappa^2+(\omega_a-\Delta)^2][\kappa^2+(\omega_a+\Delta)^2] +\gamma[(\gamma+2\kappa)\nonumber\\
&&\times(\kappa^2+\Delta^2)+2\kappa\omega^2_a]\}+\Delta\omega_aG^2(\gamma+2\kappa)^2>0 \nonumber\\
S_2&=&\omega_a(\kappa^2+\Delta^2)-G^2\Delta >0.
\end{eqnarray}
\begin{figure}
    \centering
 \subfigure[]{\includegraphics[height=1.5in, width=1.6831893in]{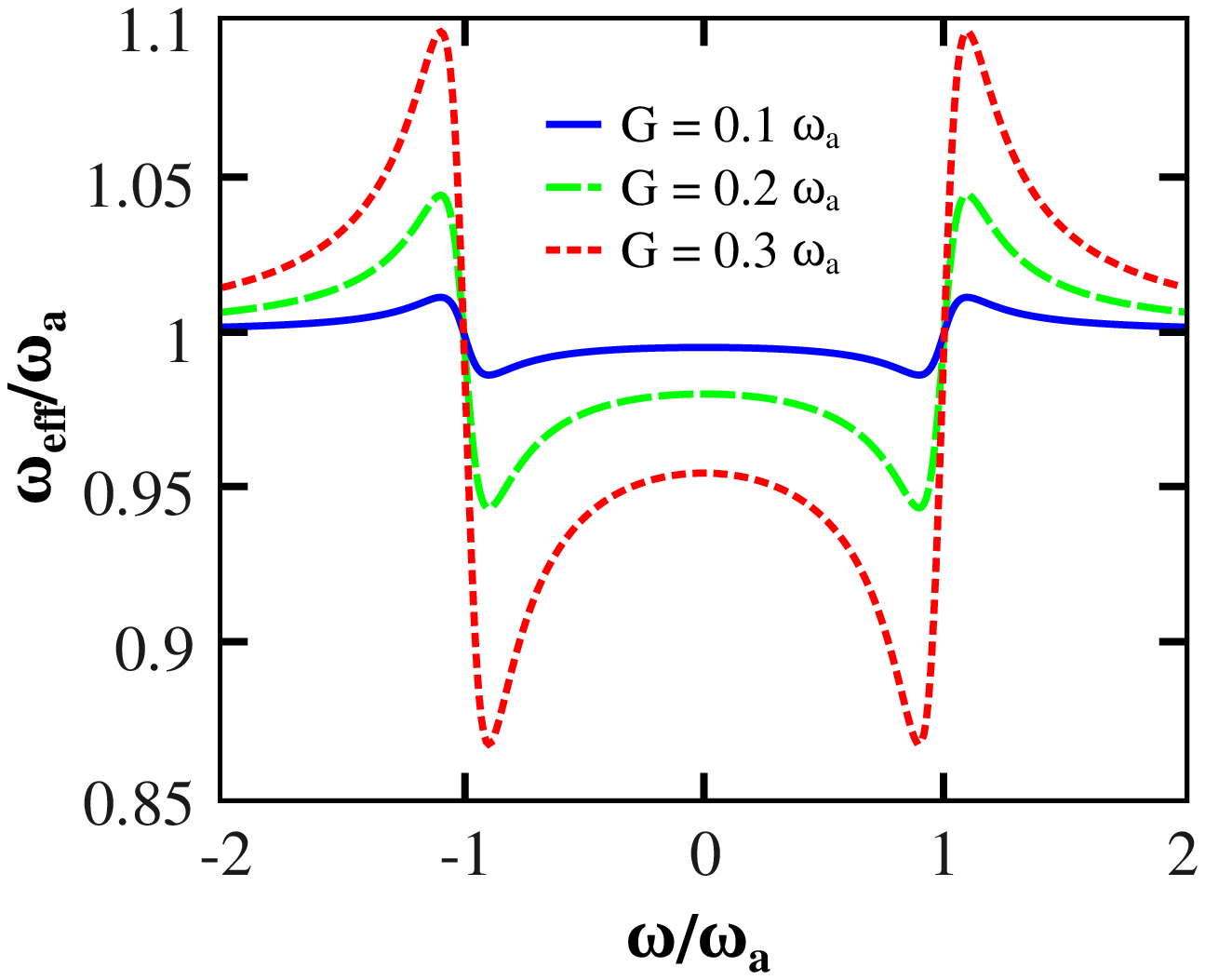}\label{weff1}}
\subfigure[]{\includegraphics[height=1.5in, width=1.6831893in]{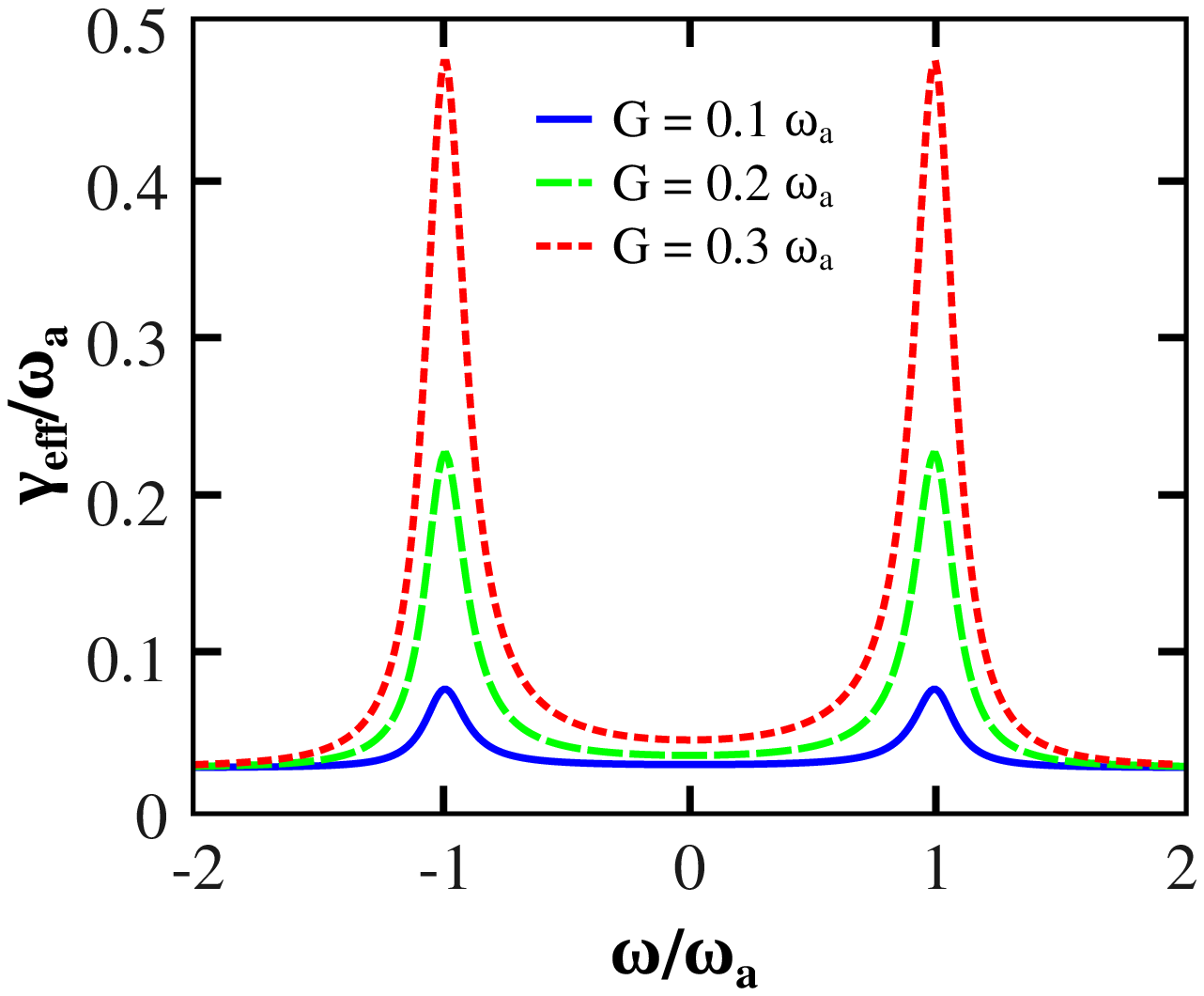}\label{geff2}}
\caption{(Color online)(a) Plot normalized effective frequency $\omega_{eff}/\omega_a$ of the condensate mode  as a function of the normalized frequency $\omega/\omega_{a}$. Parameters values are $\omega_a=2\pi\,4\times 3.8\,\rm{kHz}$, $\gamma= 2\pi\times 0.4\,\rm{kHz}$, $\Delta=\omega_a$, $\kappa=0.1\,\omega_a$ and $G=0.1\,\omega_a$ (solid blue curve), $G=0.2\,\omega_a$ (dashed green curve), and $G=0.3\,\omega_a$ (red dotted line). (b) Plot of normalized effective damping rate $\gamma_{eff}/\omega_a$ of the condensate versus normalized frequency $\omega/\omega_a$ for three different values of the BEC-field coupling,  $G=0.1\,\omega_a$ (solid blue line), $G=0.2\,\omega_a$ (dashed green line), and $G=0.3\,\omega_a$ (red dotted line).
} \label{wgeff}
\end{figure} 
\begin{figure}[b]
    \centering
\subfigure[]{\includegraphics[scale=.3355]{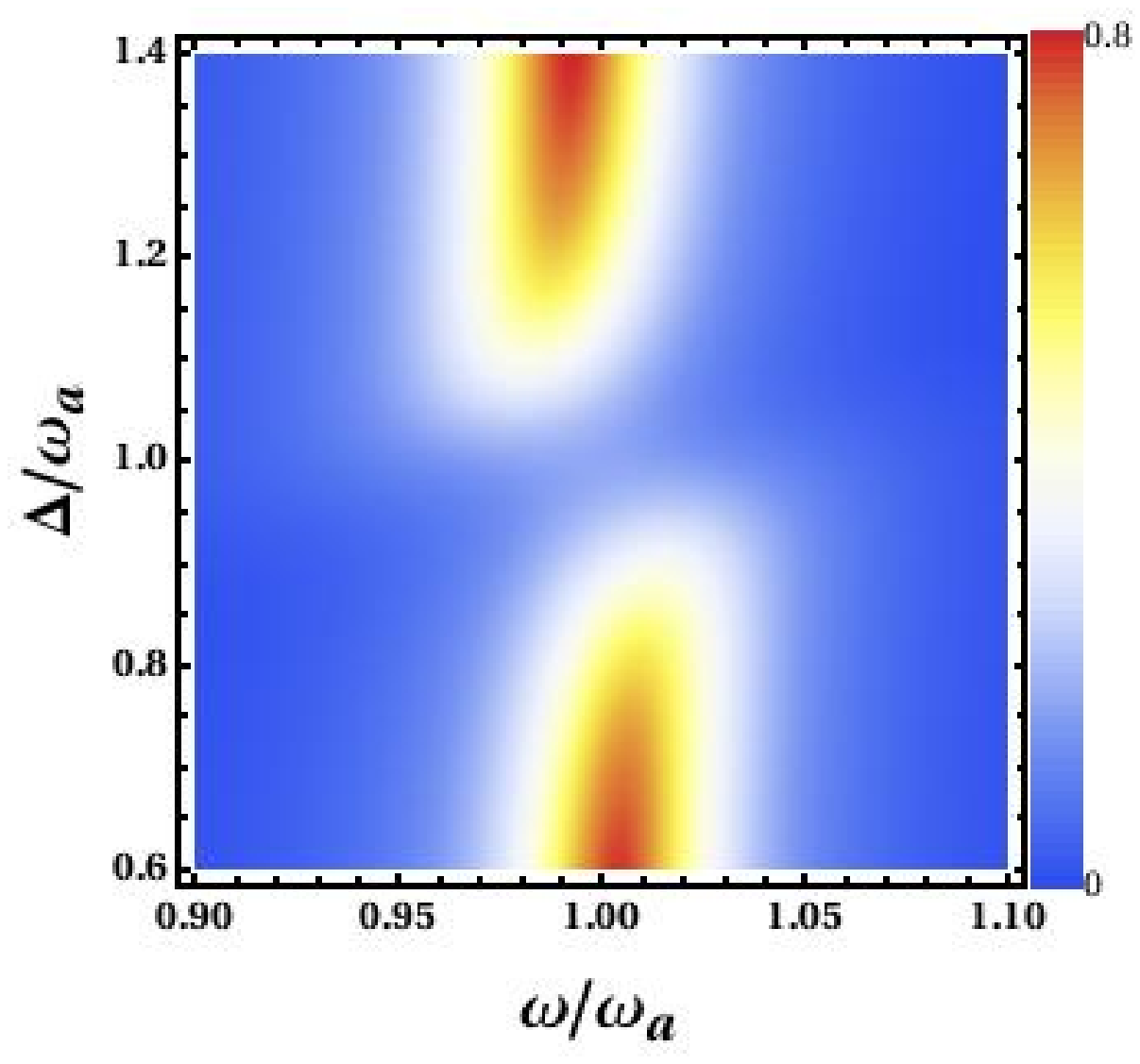}\label{s1}}
\subfigure[]{\includegraphics[scale=.3355]{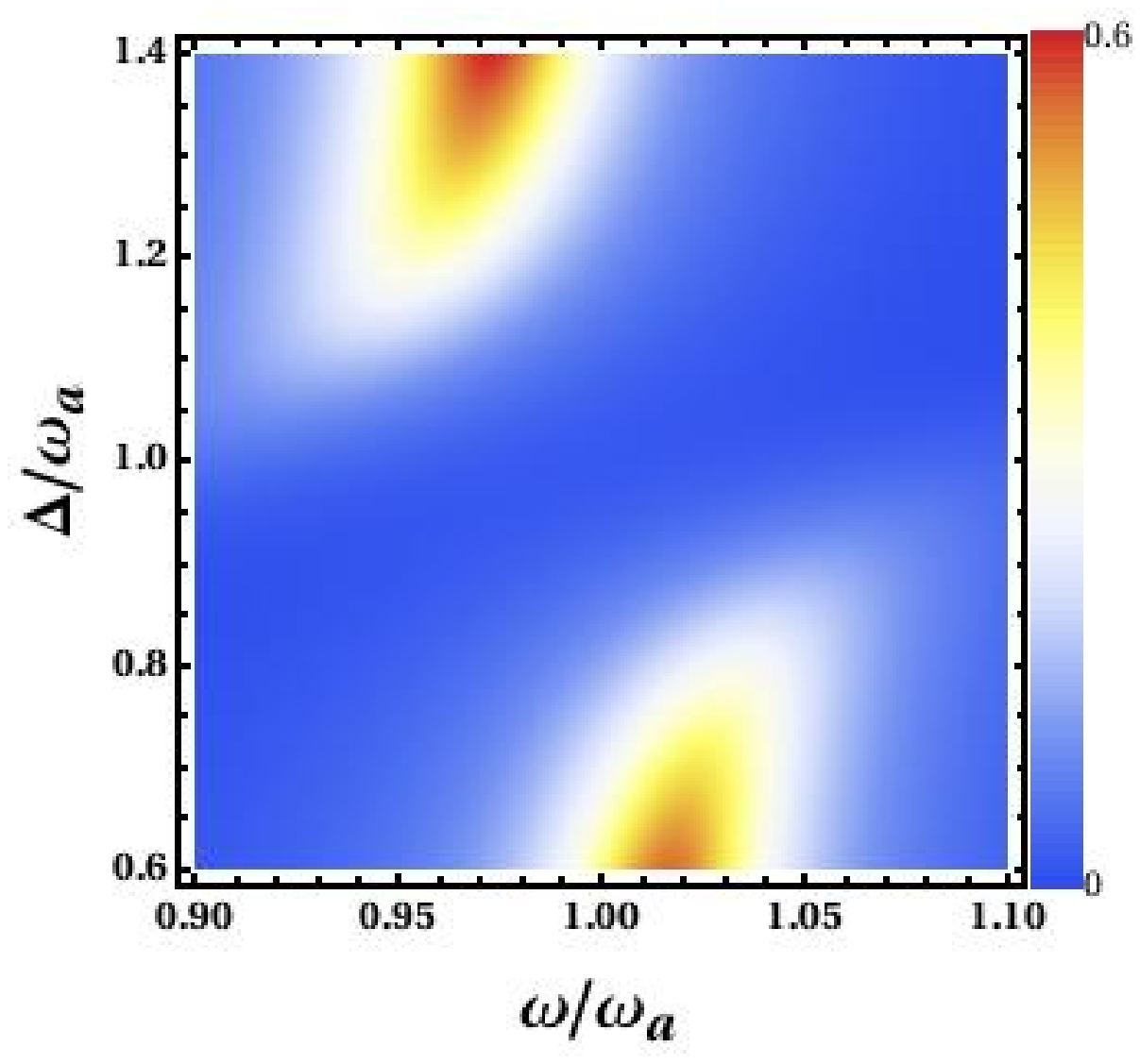}\label{s2}}
\caption{(Color online) The displacement spectrum of the Bogoliubov mode $S_{x_a}(\omega)$ as a function of the normalized detuning $\Delta/\omega_a$ and normalized frequency $\omega/\omega_{a}$ for $\kappa=0.1\,\omega_a$ and (a) $G=0.1\,\omega_a$ (b) $G=0.2\,\omega_a$. The other parameters values are the same as in Fig.\ref{wgeff}.} \label{spe}  
\end{figure}
We choose a parameter regime such that the above stability conditions are satisfied to ensure stability of the system. The Heisenberg-Langevin equations (\ref{linrz}) are linear in creation and annihilation operator. We define the quadratures, \textit{i.e}., $\partial x_a=(\partial a+\partial a^\dag)/\sqrt{2}$, $\partial p_a=(\partial a-\partial a^\dag)/i\sqrt{2}$, $\partial x_c=(\partial c+\partial c^\dag)/\sqrt{2}$ and $\partial p_c=(\partial c-\partial c^\dag)/i\sqrt{2}$, and solve the Langevin equations in Fourier space. Here, we introduce the displacement spectrum of Bogoliubov mode of collective oscillation of the BEC which is obtained from the two-frequency auto-correlation function $\langle \partial x_a(\omega)\partial x_a(\omega')\rangle=S_{x_a}(\omega)\delta(\omega+\omega')$. Therefore, the displacement spectrum  $S_{x_a}(\omega)=(1/2\pi)\int e^{-it(\omega-\omega')}\langle \partial x_a(\omega)\partial x_a(\omega')\rangle d\omega'$ of the condensate in Fourier space is given by

\begin{equation}
S_{x_a}(\omega)=\dfrac{\omega^2_a}{|d(\omega)|^2}\left[\dfrac{2G^2\kappa(\kappa^2+\omega^2_a+\Delta^2)}{(\kappa^2+\Delta^2-\omega^2)^2+4\kappa^2\omega^2}+2\gamma\right], \label{spec}
\end{equation} 
where $\omega$ stands for frequency and
\begin{equation*}
d(\omega)=\omega^2_a-\omega^2-i\omega\gamma-\dfrac{\omega_a\Delta\,G^2}{(\kappa-i\omega)^2+\Delta^2}
\end{equation*}
is the modified susceptibility of the condensate mode due to radiation pressure. The effective frequency ($\omega_{eff}$) of collective oscillation of the BEC and effective damping rate ($\gamma_{eff}$) of collective oscillation of the BEC are given by  
\begin{eqnarray}
\omega_{eff}&=& \left[ \omega^2_{a}-\dfrac{\omega_{a}\Delta G^2(\kappa^2-\omega^2+\Delta^2)}{(\kappa^2-\omega^2+\Delta^2)^2+(2\kappa\omega)^2}\right]^{\dfrac{1}{2}},\label{weff}\\
\gamma_{eff}&=&\left[\gamma+\dfrac{2\omega_a\kappa\Delta G^2}{(\kappa^2-\omega^2+\Delta^2)^2+(2\kappa\omega)^2}\right].\label{geff}
\end{eqnarray}
The frequency of the condensate mode is modified due to radiation pressure as shown in Eq.(\ref{weff}) and this is equivalent to the optical spring effect in case of opto-mechanical system with moving mirror. The spectrum of condensate mode is described by the effective susceptibility $d(\omega)$ and displacement spectrum of collective oscillation of the BEC which consists of two terms, the first term is proportional to quantum fluctuation of the radiation pressure and, second term arises from quantum noises associated with the matter waves. Therefore, the position spectrum of the condensate mode is determined by radiation pressure and quantum noise.
\paragraph*{} For our numerical calculations, we choose the parameters very close to the BEC-cavity system \cite{Ritter,Baumann}. The recoil frequency $\omega_r=2\pi\times3.8\,\rm{KHz}$, decay rate $ \gamma=2\pi\times0.4\,\rm{KHz}$, atom-field detuning $\Delta_{a}=2\pi\times 32 \,\rm{GHz}$ and single atom-photon coupling is $g_{o}=2\pi\times 10.9\,\rm{MHz}$. The optical spring effect leads to the shift in the frequency of the condensate mode and this shift is small for low-frequency oscillators \cite{Corbitt} and shift is significant for high-frequency oscillators \cite{Gigan}. In Fig.\ref{weff1}, we plot normalized effective frequency $\omega_{eff}/\omega_a$ of Bogoliubov mode of collective oscillation of the BEC as a function of the normalized frequency $\omega/\omega_{a}$ for three different values of the BEC-intracavity field interaction, $G=0.1\,\omega_a$ (solid blue curve), $G=0.2\,\omega_a$ (dashed green curve), and $G=0.3\,\omega_a$ (red dotted curve). It is noted that the deviation in the bare frequency of the condensate mode $\omega_a$ is increased as the coupling between BEC and intracavity field is increased. In Fig.\ref{geff2}, we plot the normalized effective damping rate $\gamma_{eff}/\omega_a$ of the condensate mode versus normalized frequency $\omega/\omega_a$ for different values of the BEC-field interaction. The effective damping rate of the condensate mode is significantly increased as the BEC-field interaction is increased as shown in Fig.\ref{geff2}. Therefore, strong atom-field interaction causes higher atomic loss and this atomic loss is maximum near the resonance. The back action of the light causes the heating of the atoms, and as a consequence, atomic loss was observed in \cite{Murch}. The increase in the effective damping is at the basis of cooling of Bogoliubov mode of the BEC. The BEC couples to the intracavity field  due to radiation pressure force and radiation pressure behaves as an thermal reservoir for the BEC resonator. The effective temperature of the Bogoliubov mode of the condensate takes the value between initial thermal reservoir temperature and optical reservoir. Therefore, when $G>\gamma$ one achieves the ground state cooling of the Bogoliubov mode. 
\paragraph*{}
In Fig.\ref{spe}, we plot the displacement spectrum of the BEC as a function of the normalized detuning $\Delta/\omega_a$ and normalized frequency $\omega/\omega_a$ for two different values of the BEC field interaction, \textit{i.e}, $G=0.1\,\omega_a$, Fig.\ref{s1}, and $G=0.2\,\omega_a$, Fig.\ref{s2}. For small value of interaction between condensate mode and cavity field, i.e $G=0.1\,\omega_a$, the fluctuations spectrum of the Bogoliubov mode barely shows the normal mode splitting. As we increase the coupling strength, the normal mode splitting becomes observable. For $G=0.2\omega_a$, the normal mode splitting is clearly seen in Fig.\ref{s2}. It is observed that the normal mode splits in the presence of the large coupling between BEC and field. The normal mode splitting is due to the hybridization of the two oscillators: the electromagnetic field and collective oscillations of the BEC mode. The normal modes splitting results from the mixing of the fluctuations of the cavity field with the fluctuations of the condensate. The normal mode occurs at frequencies $\omega^2_{\pm}=(\Delta^2+\omega^2_a\pm\sqrt{(\Delta^2-\omega^2_a)^2+4G^2\omega_a\Delta})/2$. In this expression, we neglect the cavity decay rate $\kappa$ and damping rate of the Bogoliubov mode. The splitting is directly proportional to the coupling rate between Bogoliubov mode and intracavity field, i.e, $\omega_{+}-\omega_{-}\propto G$.  
\paragraph*{} The atom-field system is probed by measuring the optical field transmitted by the cavity, particularly, its spectrum. The power spectrum of the light emitted by the cavity is determined from Eq.(\ref{linrz}). In frequency space, the fluctuating $\partial c(\omega)$ of the intracavity field can be obtained by using Heisenberg Langevin equations (\ref{linrz}). With the quantum input-output relation \cite{gardiner} $\partial c_{out}=\sqrt{2\kappa}\partial c(\omega)-\partial c_{in}(\omega)$, the power spectral density of the cavity output field is defined by taking the Fourier transform of its autocorrelation function $\langle c^\dag_{out}(\omega')\,c_{out}(\omega)\rangle=S_{cout}(\omega)\delta(\omega'+\omega)$ as $S_{cout}(\omega)=(1/2\pi)\int e^{-it(\omega-\omega')}\langle \partial c_{cout}(\omega)\partial c_{cout}(\omega')\rangle d\omega$. Therefore, the power spectra of the light emitted by the cavity is given by, 
\begin{equation}
 S_{cout}(\omega)=\dfrac{1}{|d_c(\omega)|^2}\left[4\kappa^2|\alpha(\omega)|^2+2\gamma|\beta(\omega)|^2\right],\label{cout}
\end{equation} 
where,
\begin{eqnarray*}
\alpha(\omega)&=& i\dfrac{\omega_aG^2}{\sqrt{2}(\omega^2_{a}-\omega^2-i\omega\gamma)},\\
\beta(\omega)&=& -i \sqrt{2\kappa}\omega_{a}G\dfrac{\kappa+i\Delta+i\omega}{\sqrt{2}(\omega^2_{a}-\omega^2-i\omega\gamma)},\\
d_c(\omega)&=& (\kappa-i\omega)^2+\Delta^2-\dfrac{\sqrt{2}\Delta G^2}{\omega^2_{a}-\omega^2-i\omega\gamma}. 
\end{eqnarray*} 
\begin{figure}[tb]
    \centering
\includegraphics[height=3in, width=3.5in]{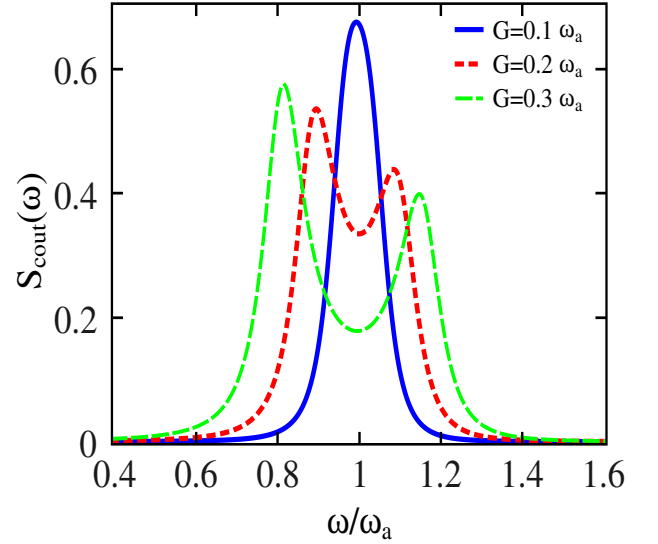}
\caption{(Color online) The emission spectrum $S_{cout}(\omega)$ as a function of the normalized frequency $\omega/\omega_a$ for different values of the BEC-field coupling strength, $G=0.1\,\omega_a$ (solid blue curve), $G=0.2\,\omega_a$ (dotted red curve) and $G=0.3\,\omega_a$ (green dashed curve). Moreover, the cavity detuning $\Delta=\omega_a$ and the other parameters are the same as in Fig.(\ref{wgeff}).}
\label{Cout1}
\end{figure} 
It is clear from Eq.(\ref{cout}) that the spectrum of the output field consists of two terms: the first term describes the coupling of intracavity field with the Bogoliubov mode of collective density oscillations of the BEC and second term accounts for the vacuum noise associated with condensate mode. 
\paragraph*{} In Fig.\ref{Cout1}, we plot the fluctuations spectrum of the output field as a function of the normalized frequency $\omega/\omega_a$ for different values of coupling strength between intracavity field with BEC. For small coupling between condensate mode and cavity field $G=0.1\, \omega_a$, one can see that only single peak (solid Blue curve) is appeared in the fluctuations spectrum of the output field. However, as the interaction of the intracavity field with BEC is $G=0.2\,\omega_a$ increased, the spectrum splits into two peaks (Red dotted curve). For $G=0.3\, \omega_a$, we observe $S_{cout}(\omega)$ further splits into two sideband peaks. This splitting is pure result of the coupling between condensate mode and cavity mode. It is also observed that the separation between two peaks in the $S_{cout}(\omega)$ is directly proportional to the BEC-field interaction strength. This splitting depends linearly on the coupling parameter $G$ and goes to zero in the absence of BEC-field coupling.
\section{conclusion}
 In conclusion, we theoretically analyze the normal mode splitting in a system which consists of Bose-Einstein condensate trapped inside a Fabry-P\'{e}rot cavity driven by laser field. We observe that the frequency and damping rate of the condensate mode display shift due to the radiation pressure force. The hybridization of the Bogoliubov mode with the fluctuations of the cavity field leads to normal mode splitting. The results show that normal mode splitting in fluctuations spectrum of the condensate mode depends linearly on the coupling of the BEC with optical field. We also analyze the coupled dynamics of the BEC and cavity field by probing the emission spectrum of the cavity. Normal mode splitting is also observed in the transmission spectra by cavity and this splitting is directly proportional to the BEC-field coupling strength. Therefore, the coupling rate between Bogoliubov mode and interactivity field can be determined by  measuring the distance between two splitting peaks.           


\begin{thebibliography}{99}
\bibitem{Kippenberg} T. J. Kippenberg and K. J. Vahala, Science \textbf{321}, 1172 (2008).
\bibitem{Braginsky}  V. B. Braginsky, \textit{Measurement of Weak Forces in Physics Experiments} (University of Chicago Press, Chicago, 1977).
\bibitem{Mancini}S. Mancini, V. Giovannetti, D. Vitali and P. Tombesi, Phys. Rev. Lett. \textbf{88}, 120401 (2002). 
\bibitem{Zhang} J. Zhang et al., Phys. Rev. A \textbf{68}, 013808 (2003).
\bibitem{Tian} L. Tian and P. Zoller, Phys. Rev. Lett. \textbf{93}, 266403 (2004).
\bibitem{Naik} A. Naik et al., Nature (London) \textbf{443}, 193 (2006).
\bibitem{Fabre} C. Fabre, M. Pinard, S. Bourzeix, A. Heidmann, E. Giacobino, and S. Reynaud, Phys. Rev. A \textbf{49}, 1337 (1994).
  \bibitem{Tombesi} S. Mancini and P. Tombesi, Phys. Rev. A \textbf{49} 4055 (1994).
 \bibitem{Caves} C. M. Caves, K. S. Thorne, R. W. P. Drever, V. D. Sandberg  and M. Zimmermann, Rev. Mod. Phys. \textbf{52} 341 (1980).
 \bibitem{LaHaye} M. D. LaHaye, O. Buu, B. Camarota and K. C. Schwab,  Science \textbf{304} 74 (2004). 
\bibitem{Rugar} D. Rugar et al., Nature (London) \textbf{430}, 329 (2004).
\bibitem{Ekinci} K. L. Ekinci , Y. T. Yang  and M. L. Roukes, J. Appl. Phys. \textbf{95} 2682 (2004).
\bibitem{Caves1980}  C. M. Caves, Phys. Rev. Lett. \textbf{45} 75 (1980).
\bibitem{V. Braginsky} V. Braginsky and S. P. Vyatchanin, Phys. Lett. A \textbf{293}, 228 (2002).
\bibitem{Marshall} W. Marshall, C. Simon, R. Penrose and D. Bouwmeester 2003 Phys. Rev. Lett. \textbf{91} 130401 (2003).
\bibitem{Paternostro} M. Paternostro, D. Vitali, S. Gigan, M. S. Kim, C. Brukner, J. Eisert, and M. Aspelmeyer, Phys. Rev. Lett. \textbf{99}, 250401 (2007).
\bibitem{Vitali} D. Vitali, S. Gigan, A. Ferreira, H. R. B\"{o}hm, P. Tombesi, A. Guerreiro, V. Vedral, A. Zeilinger, and M. Aspelmeyer, Phys. Rev. Lett. \textbf{98}, 030405 (2007).
\bibitem{Nielsen} M. A Nielsen and I. L. Chuang, \textit{Quantum Computation and Quantum Information} (Cambridge University Press, Cambridge, 2000).
\bibitem{Ian} H. Ian, Z. R. Gong, Y. X. Liu, C. P. Sun, and F. Nori, Phys. Rev. A \textbf{78}, 013824 (2008).
\bibitem{genes} C. Genes, D. Vitali, and P. Tombesi, Phys. Rev. A \textbf{77}, 050307(R) (2008).
\bibitem{Meiser} D. Meiser and P. Meystre, Phys. Rev. A \textbf{73}, 033417 (2006).
\bibitem{Ritsch} C. Genes, H. Ritsch, and D. Vitali, Phys. Rev. A \textbf{80}, 061803(R) (2009).
\bibitem{Thompson} R. J. Thompson, G. Rempe, and H. J. Kimble, Phys. Rev.
Lett. \textbf{68}, 1132 (1992).
\bibitem{Wallraff} A. Wallraff et al., Nature (London) \textbf{431}, 162 (2004).
\bibitem{Reithmaier} J. P. Reithmaier et al., Nature (London) \textbf{432}, 197 (2004);
T. Yoshie et al., ibid. \textbf{432}, 200 (2004); K. Hennessy et al.,
ibid. \textbf{445}, 896 (2007).

\bibitem{Esteve} J. Esteve et al., Nature (London) \textbf{455}, 1216 (2008).
\bibitem{Brennecke}  F. Brennecke, T. Donner, S. Ritter, T. Bourdel, M. K\"{o}hl, and T. Esslinger, Nature (London)  \textbf{450}, 268 (2007).
\bibitem{Ritter} F. Brennecke, S. Ritter, T. Donner, and T. Esslinger, Science \textbf{322}, 235 (2008).
\bibitem{Colombe}Y. Colombe, T. Steinmetz, G. Dubois, F. Linke, D. Hunger, and J. Reichel, Nature (London) \textbf{450}, 272 (2007).
\bibitem{Purdy} T. P. Purdy, D. W. C. Brooks, T. Botter, N. Brahms, Z. Y. Ma, and D. M. Stamper-Kurn, Phys. Rev. Lett. \textbf{105}, 133602 (2010).
\bibitem{Baumann} S. Ritter, F. Brennecke, K. Baumann, T. Donner, C. Guerlin, and T. Esslinger, Appl. Phys. B: Lasers Opt. \textbf{95}, 213 (2009).
\bibitem{Kónya} D. Nagy, G. K\'{o}nya, G. Szirmai, and P. Domokos, Phys. Rev. Lett. \textbf{104}, 130401 (2010).
\bibitem{Nagy} D. Nagy, P. Domokos, A. Vukics, and H. Ritsch, Eur. Phys. J. D \textbf{55}, 659 (2009).
\bibitem{asjad} Muhammad Asjad and Farhan Saif, Phys. Rev. A \textbf{84}, 033606 (2011).
\bibitem{cw} C. W. Gardiner, Phys. Rev. Lett. \textbf{56}, 1917 (1986).
\bibitem{Hurwitz} A. Hurwitz, in \textit{Selected Papers on Mathematical Trends in Control Theory}, edited R. Bellman and R. Kalaba (New York, Dover, 1964). 
\bibitem{Corbitt} T. Corbitt, Y. Chen, E. Innerhofer, H. M\"{u}ller-Ebhardt, D. Ottaway, H. Rehbein, D. Sigg, S. Whitcomb, C. Wipf, and N. Mavalvala, Phys. Rev. Lett. \textbf{98}, 150802 (2007); T. Corbitt, C. Wipf, T. Bodiya, D. Ottaway, D. Sigg, N. Smith, S. Whitcomb, and N. Mavalvala, ibid.  \textbf{99}, 160801 (2007).
\bibitem{Gigan} S. Gigan, H. B\"{o}hm, M. Paternostro, F. Blaser, G. Langer, J. Hertzberg, K. Schwab, D. B\"{a}uerle, M. Aspelmeyer, and A. Zeilinger, Nature (London)  \textbf{444}, 67 (2006); O. Arcizet, P.-F. Cohadon, T. Briant, M. Pinard, and A. Heidmann, Nature (London)  \textbf{444}, 71 (2006); A. Schliesser, P. Del’Haye, N. Nooshi, K. J. Vahala, and T. J. Kippenberg, Phys. Rev. Lett. 97, 243905 (2006).
\bibitem{Murch}  K. W. Murch et al Nature Phys. \textbf{4} 461 (2008). 
\bibitem{gardiner}  C. W. Gardiner, \textit{Quantum Noise} (Berlin: Springer 1991).
 \end{thebibliography}
\end{document}